\documentclass[aps,prl,superscriptaddress,twocolumn,floatfix]{revtex4-1}

\usepackage{graphicx,graphics}

\usepackage{dcolumn}

\usepackage{latexsym,verbatim}

\usepackage{bm}

\usepackage{color}

\usepackage[percent]{overpic}

\usepackage[breaklinks=true,colorlinks,citecolor=blue,linkcolor=blue,urlcolor=blue]{hyperref}

\usepackage{float}

\begin{document}

\author{Trond I. Andersen}
\affiliation{Department of Physics, Harvard University, Cambridge, MA 02138, USA}

\author{Thomas B. Smith}
\affiliation{School of Physics and Astronomy, University of Manchester, Manchester M13 9PL, United Kingdom}

\author{Alessandro Principi}
\affiliation{School of Physics and Astronomy, University of Manchester, Manchester M13 9PL, United Kingdom}

\title{Enhanced photoenergy harvesting and extreme Thomson effect in hydrodynamic electronic systems}

\begin{abstract}
The thermoelectric (TE) properties of a material are dramatically altered when electron-electron interactions become the dominant scattering mechanism. In the degenerate hydrodynamic regime, the thermal conductivity is reduced and becomes a {\it decreasing} function of the electronic temperature, due to a violation of the Wiedemann-Franz (WF) law. We here show how this peculiar temperature dependence gives rise to new striking TE phenomena. These include an 80-fold increase in TE efficiency compared to the WF regime, dramatic qualitative changes in the steady state temperature profile, and an anomalously large Thomson effect. In graphene, which we pay special attention to here, these effects are further amplified due to a doubling of the thermopower.

\end{abstract}

\maketitle

{\it Introduction.---} For decades, the holy grail of thermoelectricity (TE) has been the enhancement of the heat-to-work conversion efficiency of TE devices to the ultimate limit allowed by thermodynamics, {\it i.e.} the Carnot efficiency $\eta_{\rm C}$~\cite{Sherman1960}.
The degree to which a TE system approaches the Carnot limit increases with the (dimensionless) figure-of-merit $zT \equiv \sigma S \Pi/\kappa$~\cite{Kim2015,Sherman1960,Nemir2010}, where $\sigma$ and $\kappa$ are the electrical and thermal conductivities, $S$ and $\Pi = T S$ are the Seebeck (or thermopower) and Peltier coefficients~\cite{Ashcroft}, and $T$ is the electronic temperature. 
Intuitively, lowering $\kappa$ allows a system to sustain higher temperature gradients, and $S$ determines the amount of electricity that can be generated from these. Finally, high values of $\sigma$ minimize the energy lost in the conversion process. There have been numerous efforts to improve $zT$~\cite{Hsu2004,He2017}, focusing on both increasing the Seebeck coefficient~\cite{Xu2014, Lee2006, Kim2006, Takahashi2016, Tan2016}, and reducing the thermal conductivity~\cite{Hu2014,Morelli2008,Fu2015,Zhao2014,Lehmann2015,Carrete2014,Lee2014,Yu2010}. 
The latter task is especially nontrivial, because even if the phononic contribution to $\kappa$ is minimized, its electronic part
$\kappa_{\rm{e}}$
 is connected to $\sigma$ at a fundamental level~\cite{Wiedemann_Franz,Ashcroft}. When charge and heat currents are carried by the same quasiparticles ({\it i.e.} the electrons), they are impeded by scattering against disorder to the same extent~\cite{Castellani_prl_1987,Arfi1992,Schwab_adp_2003,Raimondi_prb_2004,Catelani2005}. This gives rise to the proportionality between the respective conductivities, known as the Wiedemann-Franz (WF) law: $\kappa_{\rm WF}=\mathcal{L}_0\sigma T$. Here, $\mathcal{L}_0=\pi^2k_B^2/(3e^2)=2.44\cdotp10^{-8}~{\rm W}\Omega \mathrm{K}^{-2}$ is the Lorenz number, a universal constant of nature, $k_{\rm B}$ the Boltzmann constant, and $e$ the electronic charge.

In the Fermi-liquid regime ($k_{\rm B} T \ll \mu$), the thermopower $S$ is given by $S_{\rm WF} =-\pi^2 k_{\rm B}^2 T/(3 e) \partial\ln \sigma/\partial\mu\to-\beta\pi^2 k_{\rm B}^2 T/(3 e \mu)$ for $\sigma\propto\mu^{\beta}$. At best, when the phononic contribution to the thermal conductivity vanishes, one finds $zT_{\rm WF} = \pi^2 \beta^2 (k_{\rm B} T)^2/(3 \mu^2)$, which is small and without much room for improvement for the given $\sigma(\mu)$. 

The WF proportionality breaks down when electron-electron (e-e) collisions are the dominant scattering mechanism~\cite{Principi2015,Xie2016,Gooth_arxiv_2017}, often referred to as the ``hydrodynamic'' regime ~\cite{Andreev2011,Jong1995,Lucas2018,Mendoza2013,Moll2016,Kumar2017,Kumar2017a,Bandurin2016}.
Crucially, and in contrast to impurity scattering, e-e interactions affect charge and heat currents very differently.
While momentum-conservation entails that charge currents are unaffected by e-e collisions, heat currents are not conserved. The thermal conductivity is therefore reduced, according to~\cite{Principi2015}
\begin{equation} \label{eq:kappadef}
\kappa_{\rm Hyd} = \frac{\kappa_{\rm WF}}{1+8\Gamma_{\rm ee}/(5\Gamma_{\rm mr})}=\frac{\mathcal{L}_0\sigma T}{1+(T/T_{\rm int})^2}
~,
\end{equation}
where $\Gamma_{\rm ee}$ and $\Gamma_{\rm mr}$ are the rates of e-e and momentum-relaxing scattering processes, respectively. This violation of the WF law proves very advantageous from a TE perspective, as we show below.

In graphene, which we will focus on here, $\Gamma_{\rm mr} = e^2\mu/(\pi \hbar^2 \sigma)$ and $\Gamma_{\rm ee}\simeq\pi k_B^2T^2/(4\hbar\mu)$~\cite{Kotov2012,polini2016}, which allows us to define $T_{\rm int}=\sqrt{5\hbar \Gamma_{\rm mr} \mu/(2\pi k_B^2)}$ as the characteristic electronic temperature at which hydrodynamic effects become important. We then find $T_{\rm int}\sim48.5 ~{\rm K}$ at $\mu=100~{\rm meV}$ with a realistic mean free path of $L_{\rm mfp}=3~\mu{\rm m}$~\footnotetext{Strong e-e interactions ensure that the carriers are thermalized locally and prevent ballistic behavior, despite the long (momentum-relaxing) $L_{\rm mfp}$}\cite{Note1,Mayorov2011,Wang2013,Kumar2017,Bandurin2016}. 
Eq.~(\ref{eq:kappadef}) has two striking consequences. First, $\kappa_{\rm{e}}$ can be dramatically reduced in the hydrodynamic regime, reaching an impressive 39-fold reduction (compared to the WF regime) if the electrons are heated to 300 K while keeping the lattice cold. Second, and even more surprisingly, $\kappa_{\rm{e}}$ {\it decreases} with electronic temperature, and displays a rare $1/T$-dependence for $T\gtrsim T_{\rm int}$. When a region heats up, e-e collisions become more frequent, and the ability to cool by conduction decreases.
This is in stark contrast to the WF regime (where $\kappa_{\rm{e}}\propto T$), and gives hydrodynamic systems a much stronger ability to focus heat into hot spots.
Moreover, in the hydrodynamic regime, the thermopower of graphene coincides with the entropy density, causing the former to double in value ($S_{\rm Hyd}=2S_{\rm WF}$)~\cite{Muller_prb_2008,Foster_prb_2009}.

In passing, we note that the behavior exhibited by the thermal conductivity in Eq.~(\ref{eq:kappadef}) is opposite to the one reported in Ref.~\onlinecite{Crossno2016} for graphene at charge neutrality.
In the Dirac fluid ($k_{\rm B} T \gg \mu$), the coexisting electrons and holes move in the same (opposite) directions to carry heat (charge) currents. Thus, in that regime, electron-hole collisions impede charge currents, but not heat currents. 

In this Letter, we show that the e-e interactions dramatically enhance the TE efficiency, and give rise to qualitatively different temperature profiles. Whereas a lot of effort has been put into engineering optimal spatial profiles of $S$~\cite{Cui2003,Huang2013,Dashevsky2002,Kuznetsov2002}, we here shed light on the great potential of the much less considered $\kappa_{\rm{e}}(T)$ (effectively $\kappa_{\rm{e}}(x)$). Moreover, while desired spatial profiles are often achieved by designing composite materials, the efficiency enhancement shown here is due to the intrinsic $\kappa_{\rm{e}}(T)$ of a single material. In the same intrinsic material (and regime), the efficiency is increased further by the doubled $S$, and the bottlenecked phononic heat transport due to weak e-ph coupling\footnotetext{
When heat is initially injected into the electronic system, it must be transferred to the phonons before phononic heat transport can occur. Weak e-ph coupling bottlenecks this process in short devices.}\footnotetext{See Supplementary Material at [URL] for further explanation of the weak role of phonons and dependence on device length.}~\cite{Note2,Note3}.

We consider the unusual TE behavior in two scenarios. Although we focus on high-quality graphene as our model system, most conclusions are general and apply to other degenerate systems in the hydrodynamic regime as well~\cite{Principi2015,Gooth_arxiv_2017,Fu_arxiv_2018,Moll2016}.

First, we consider a photoenergy harvesting scenario, where electricity is generated from light shone on a p-n junction through the photothermoelectric (PTE) effect~\cite{Gabor2011,Ma2014,Song2011}. 
In stark contrast to the WF regime, e-e interactions give rise to a {\it convex} temperature profile, whose amplitude grows {\it superlinearly} with the incident power. These are signatures of the system's capability to retain heat in hot spots, which, combined with the doubled $S$, leads to efficiencies up to 80 times larger than in the WF regime. 

Next, we consider a graphene channel that is Joule-heated by a current injected through thermally anchored contacts. Due to a combination of the Seebeck and Peltier effects (explained below), the temperature profile is skewed in the direction of the particle flow~\cite{Bakan2014,Jungen2006,Grosse2014}. Commonly referred to as the ``Thomson effect'', this phenomenon strongly depends on the local value of $zT$ and is typically weak in conventional systems. In the hydrodynamic regime, however, we show that the temperature dependence of $\kappa_{\rm Hyd}$ and the increased $S$ drastically amplify the Thomson effect. The temperature peak shifts up to 50$\%$ of the way to the contact, making this phenomenon a potential experimental signature of hydrodynamic heat transport. In both scenarios, we also present results where phonon-polaritons are included as an extrinsic cooling mechanism~\cite{Principi2017,Tielrooij2017,Yang2017}, to show that our predictions should be observable under realistic conditions. 

{\it The theoretical model.---}We consider an hBN-encapsulated graphene sheet of length $L$ and width $W$, heated by either photoexcitation or electrical current. We assume $W$ to be small compared to both the laser spot and $L$, so that the problem is effectively one-dimensional in both scenarios. The electrons conduct heat to contacts on both sides of the device, which are thermally anchored at $50$ K. In the PTE scenario, a split gate is used to form a p-n junction, while in the Joule-heated case the charge density is uniform. In the steady state, the heat equation is 
\begin{equation} \label{eq:heat_1}
\partial_x (\kappa \partial_xT) = - \sigma^{-1}j^2 - p_{\rm in} + p_{\rm out} + \left( \partial_T \Pi - S\right) j\partial_x T
~,
\end{equation}
where $j$ is the homogeneous charge current, and $p_{\rm in}\equiv p_{\rm in}(x)$ and $p_{\rm out}\equiv p_{\rm out}(x)$ are the laser intensity and phonon cooling power density, respectively. The first term on the right represents Joule heating, while the left side describes the diffusion of heat towards the contacts.
Finally, the last term stems from the combination of Seebeck and Peltier effects. 
Since $\Pi = T S$ and $S\propto T$ in the 
Fermi-liquid regime, the round bracket in Eq.~(\ref{eq:heat_1}) equals $(+)S$. This is the so-called Thomson term that gives rise to asymmetric temperature profiles.

All TE coefficients depend on the local value of only the temperature, since the density is kept fixed by the gate. To simplify our analysis, we neglect deviations of $\mu$ from its zero-temperature value, since these are exponentially suppressed in the degenerate regime.

To be solved, Eq.~(\ref{eq:heat_1}) requires the knowledge of not only TE coefficients, but also the cooling pathways contributing to $p_{\rm out}$. Heat transfer to the graphene lattice is highly inefficient in ultra-clean graphene devices~\cite{Tse2009,Song2011,Song2015}. Direct acoustic-phonon cooling is limited by the mismatch between the Fermi and sound velocities~\cite{Bistritzer2009}, and the optical phonon energy is too high ($\sim 200$ $\rm meV$, 2400 K)~\cite{Piscanec2004} to allow for efficient coupling. Moreover, the low impurity density strongly suppresses disorder-assisted (supercollision) processes~\cite{Song2012}. The weak coupling allows for heating the electrons out of equilibrium with the lattice~\cite{Yamoah2017,Guo2018}, thus increasing the e-e scattering rate $\Gamma_{\rm ee}\propto T^2$, while only minorly affecting $\Gamma_{\rm mr}$ for low heating powers. To a good approximation, we can therefore assume that we remain in the Ohmic regime for the powers considered here \footnote{A weak electronic temperature dependence, $\Gamma_{\rm mr}(T)\simeq \Gamma_{\rm mr}(0)+B k_{\rm B}T/\mu$, would just introduce a (negative) subleading term of order $T^{-2}$ in $\kappa_{\rm{e}}$}. Utilizing this separation of electronic and phononic temperatures carries great potential for enhancing hydrodynamic effects.

Since heat transfer to the graphene lattice is so slow in high-quality 2D heterostructures, the phonon-polaritons of the hBN encapsulant~\cite{Jacob2014,Dai2015} have been shown to represent the main cooling pathway~\cite{Principi2017,Tielrooij2017,Yang2017}. These Fabry-Perot-like modes, propagating in the ``cavity'' formed by the hBN slabs, cluster around $100$ and $200~{\rm meV}$ (the so-called ``{\it Reststrahlen} bands''). Due to their large density of states, these modes can extract heat more effectively than the graphene phonons. In order to be consistent with today's state-of-the-art graphene devices, we include this cooling pathway in our simulations. This allows us to confirm that electronic heat conduction dominates over phononic contributions. At the electronic temperatures reached here, the coupling to phonon-polaritons is still weak enough to bottleneck phononic heat transport~\cite{Note3}.

{\it Enhanced photoenergy harvesting.---}We consider the photovoltage produced by a laser-heated p-n junction located in the middle of the channel, at $x=0$. The laser excites electrons from below to well above the Fermi surface, which quickly relax through rapid e-e interactions (10-100 fs) to a thermal distribution characterized by a (non-uniform) temperature profile~\cite{Tomadin2013,Brida2013,Tielrooij2013}. The subsequent heat transport is described by Eq.~(\ref{eq:heat_1}), solved with an energy source located at the junction, $p_{\rm in}=P/W\delta(x)$.

\begin{figure}[t]

\begin{center}

\begin{overpic}[width=0.9\columnwidth]{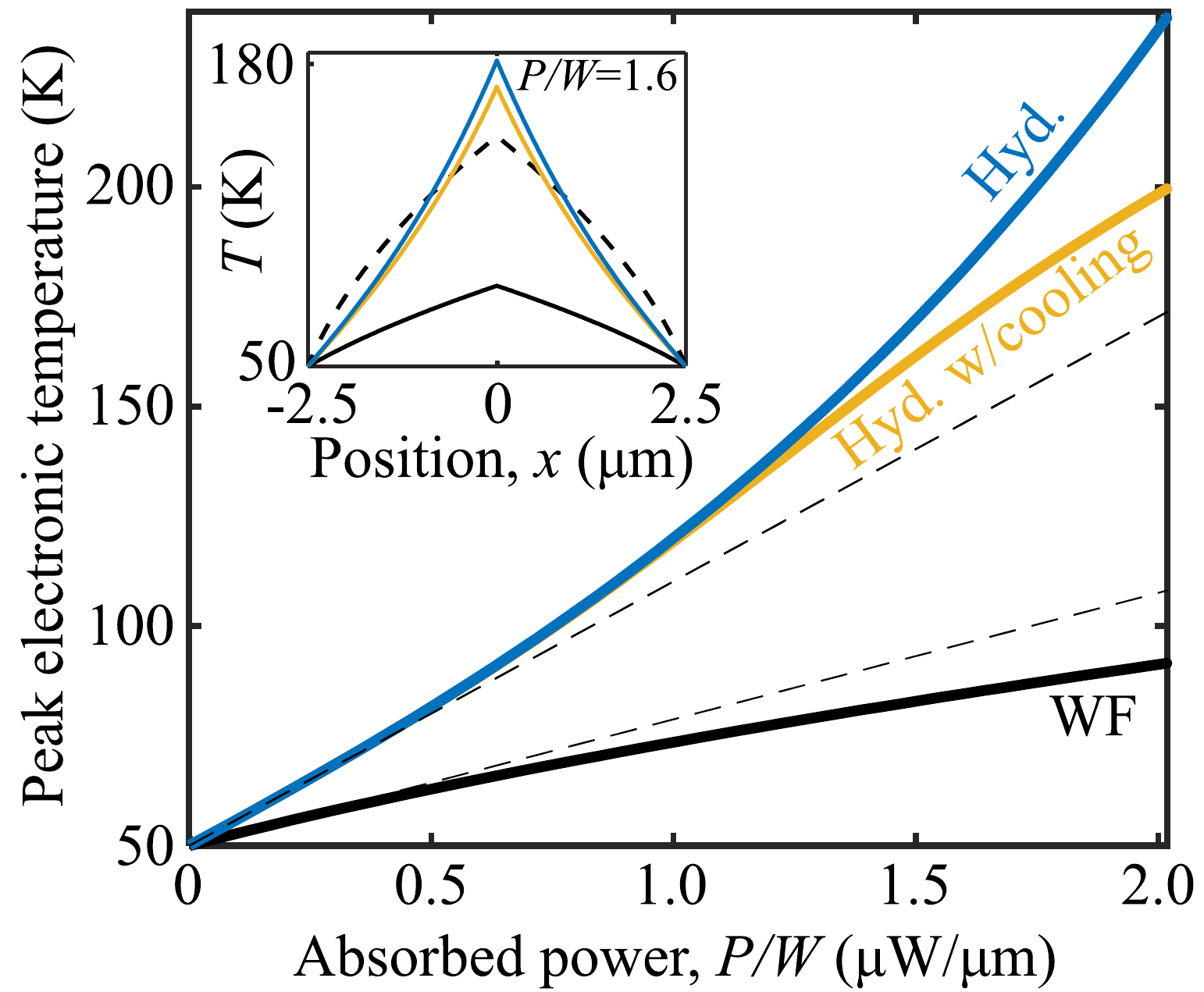}

\end{overpic}

\end{center}

\caption{(Color online)
The electronic temperature increases superlinearly with laser intensity in the hydrodynamic regime (blue), in contrast to the WF regime (black). The yellow curve includes phonon-polariton cooling. Dashed lines are guides to the eye.
Inset: temperature profiles in the same three cases. The dashed line shows the WF regime with a reduced thermal conductivity (see main text). 
}

\label{fig:one}

\end{figure}

Although a resistor must be connected to generate electricity from such a system, we first consider the open-circuit case to shed light on the role of e-e interactions. When light is shone on the p-n junction, a symmetric, peaked temperature profile forms, and a Seebeck voltage is generated across the device according to $V_{\rm PTE} = \int_{-L/2}^{L/2} S(x) \partial_x T(x) dx$. Despite the symmetry of the temperature profile, $V_{\rm PTE}$ is non-zero due to the opposite signs of $S$ in the p- and n-doped regions. In the absence of interactions, the temperature profile is everywhere concave
\begin{eqnarray} \label{Seqn1}
T_{\rm WF}(x) = T_{\rm c}\sqrt{1+2\theta f_{\rm PTE}(2x/L)}
~,
\end{eqnarray}
When e-e collisions dominate, however, the temperature profile becomes convex on each side of the pn-junction:
\begin{eqnarray} \label{Seqn2}
T_{\rm Hyd}(x) = T_{\rm c}\sqrt{\left(1+\gamma^{-1}\right)\exp\left[2\theta \gamma f_{\rm PTE}(2x/L)\right]-\gamma^{-1}}
~.
\end{eqnarray}
Here $\gamma=(T_{\rm{c}}/T_{\rm{int}})^2$, $f_{\rm PTE}(u) = 1-|u|$, and $\theta = P/P_{\rm c}$, where $P_{\rm c}=4\sigma\mathcal{L}_0 T_{\rm c}^2 W/L$ represents the characteristic cooling rate due to heat conduction towards the contacts.
In the inset of Fig.~\ref{fig:one} we show the temperature profiles in the WF (black) and hydrodynamic (blue) regimes, as well as the latter case with phonon-polariton cooling (yellow). Here, and in what follows, we set $L=5$ $\mu\rm m$, $W=500$ $\rm nm$, $|\mu|=0.1$ $\rm eV$ and $L_{\rm mfp}=3$ $\mu\rm m$ \cite{Note1}.


Qualitative and quantitative differences between the two regimes are immediately evident. While convex regions are not observed in the WF regime, the strong e-e interactions allow for these, since hotter regions need higher gradients to conduct the same heat. The enhanced focusing of heat at the junction is highly beneficial for PTE energy conversion. This allows for heating the device center to a high temperature, and also keeping low levels of phonon cooling (due to low $T$) away from the temperature peak. We stress that the convex regions are due to the uncommon temperature dependence of $\kappa_{\rm{e}}$, rather than its overall reduction. To highlight this, we also plot results for the WF case with a reduced $\kappa_{\rm WF}$, such that the average temperature is the same as in the hydrodynamic regime (dashed black curve). This is clearly not sufficient to achieve the same level of heat concentration.
We would like to note that the $\textit{phononic}$ $\kappa_{\rm{ph}}$ can show a $1/T$-dependence at high temperatures~\cite{Landau10}. However, this would only affect the electronic $T(x)$ if heat loss to the phonons played a dominant role. This is shown not to be the case here, and is generally atypical of hydrodynamic electronic systems. In terms of TE efficiency, phonon cooling is undesirable, since only electrons can convert heat to electrical current.

The main panel shows that the peak electronic temperature increases {\it superlinearly} with $P$ in the hydrodynamic regime, in stark contrast to what is observed in the WF regime, as well as in phonon-limited cases. This highly rare effect is due to the anomalous temperature dependence of $\kappa_{\rm{e}}$, which makes it progressively easier to heat the sample as the electronic temperature increases. Johnson noise thermometry~\cite{Crossno2016} can be used to probe this new signature of hydrodynamics.

We now close the circuit with an external resistor of optimal resistance $R_{\rm{opt}}$ and allow a current $I_{\rm PTE}$ to flow, to calculate the TE efficiency $\eta=I_{\rm PTE}^2R_{\rm{opt}}/P$. Although representative, the analytical $T(x)$ obtained in Eqs.~(\ref{Seqn1}-\ref{Seqn2}) are no longer exact, since $I_{\rm PTE}$ lowers the temperature through Peltier cooling. Intuitively, heat is drawn from the device to power the resistor. In the low-power regime, this effect is small, so $R_{\rm{opt}}$ is the device resistance $L/(W\sigma)$, and: 

\begin{equation} \label{eq:eta}
\eta_{\rm{Hyd}}=\frac{\pi^2}{12\theta}\left[\frac{k_{\rm{B}}T_{\rm{c}}}{\mu}\left(1+\gamma^{-1}\right)\left(e^{2\theta\gamma}-1\right)\right]^2
~.
\end{equation} 

\begin{figure}[t]

\begin{center}

\begin{overpic}[width=3.07in]{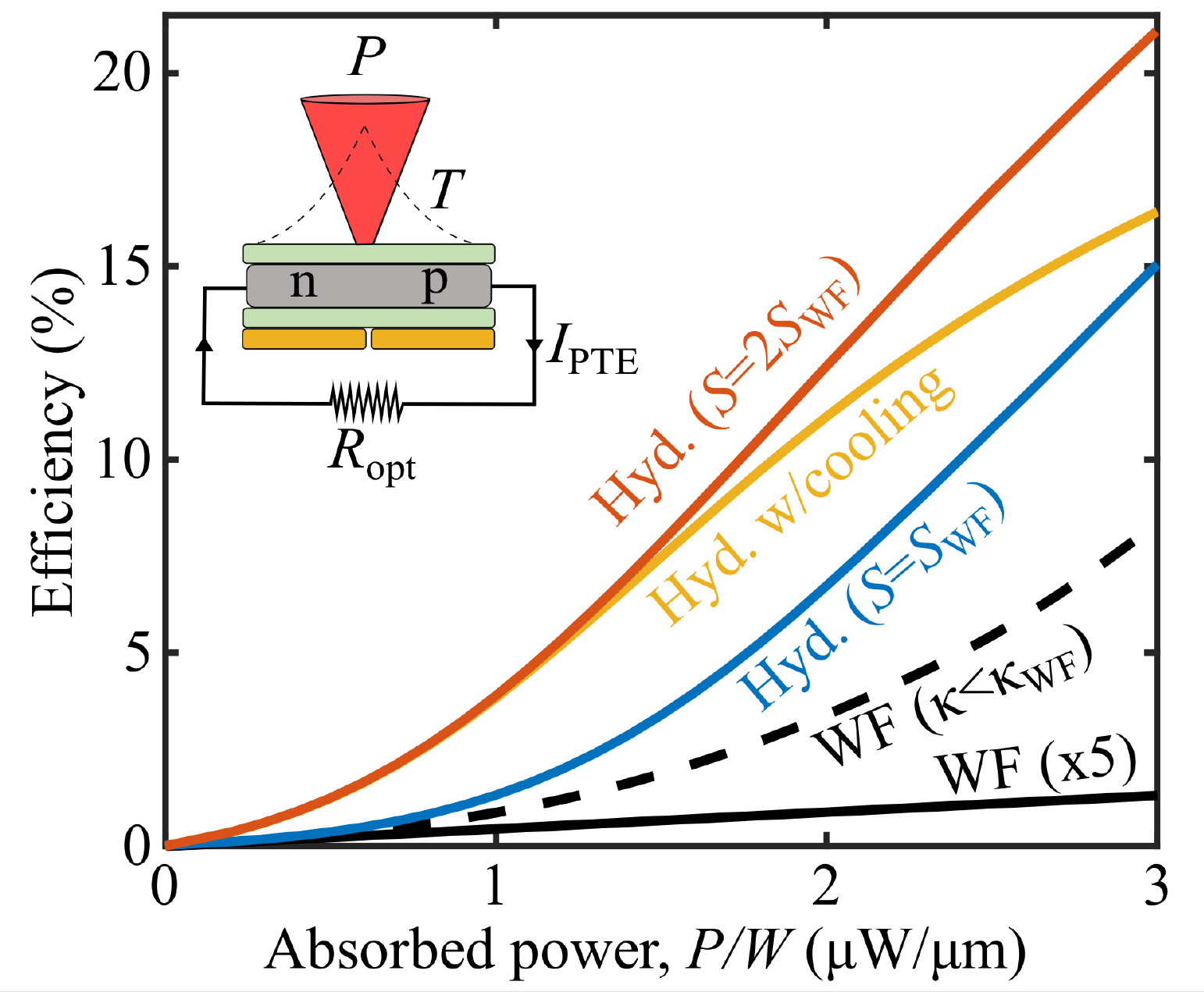} 

\end{overpic}

\end{center}

\caption{(Color online)
The hydrodynamic regime (colored) exhibits a much higher TE efficiency than the WF regime (black solid), here plotted against the incident laser heating intensity. 
We also plot the WF result with a reduced thermal conductivity (black dashed), as in Fig.~\ref{fig:one}. ($L=5$ $\mu\rm m$). 
}

\label{fig:two}

\end{figure}

Fig.~\ref{fig:two} shows the (numerically evaluated) efficiency for a larger range of powers. We observe that the efficiency becomes a striking 80 times larger in the hydrodynamic regime than in the WF case. With a peak temperature of $230$ K at $P/W=3$ $\mu\rm{W}/\mu\rm{m}$, the hydrodynamic case reaches an impressive $27\%$ of the Carnot efficiency ($\eta_{\rm{C}}=78\%$).

To separate the various effects that play a role in this dramatic efficiency improvement, we show multiple different curves. For the hydrodynamic case, we plot the efficiency for both $S=S_{\rm WF}$ and $S=S_{\rm Hyd}=2S_{\rm WF}$, to display the effect of the doubled Seebeck coefficient. In the low-$P$ regime, where the p-n junction is mainly cooled by conduction, doubling $S$ gives a four-fold increase in efficiency, since $\eta\propto V_{\rm PTE}^2$. At higher powers, Peltier cooling becomes more important, so the gain from the doubled $S$ decreases. To also show the significance of the $1/T$-dependence of $\kappa_{\rm{e}}$, we again plot the WF case with a reduced $\kappa_{\rm WF}$, such that the average temperature is equal to that of the hydrodynamic case (with $S=S_{\rm WF}$) for each incident power. Evidently, this is not enough to explain the efficiency enhancement. 


{\it Joule-heating scenario.---}We now turn to the case of a graphene channel heated by a current injected through the contacts. To clarify the role of various thermoelectric mechanisms, we will first consider the low-bias case, where the Seebeck and Peltier terms in Eq.~(\ref{eq:heat_1}) can be neglected. Then the temperature profiles can be written on the forms shown in Eqs. ($\ref{Seqn1}$-$\ref{Seqn2}$), but now with $f_{\rm PTE}$ replaced with $f_{\rm J}(u)=(1-u^2)/2$. The heating power in $\theta=P_{\rm J}/P_{\rm c}$ is now the Joule power $P_{\rm J} = \sigma V^2 W/L$, where $V= j L/\sigma$ is the voltage applied across the slab. Thus, $T_{\rm{Hyd}}(x)$ is Gaussian in the limit $\gamma\gg1$. The characteristic width of the temperature peak is $L\gamma\theta$ which, quite strikingly, decreases with Joule power. This is because the ability to focus heat is enhanced as electrons are heated up and scatter more often with each other. When $\theta\gamma \gg 1$, the interactions produce a sharply peaked temperature profile, with convex regions on the sides, in stark contrast to the approximately parabolic $T_{\rm WF}(x)$. 

\begin{figure}[t]

\begin{center}

\begin{overpic}[width=0.9\columnwidth]{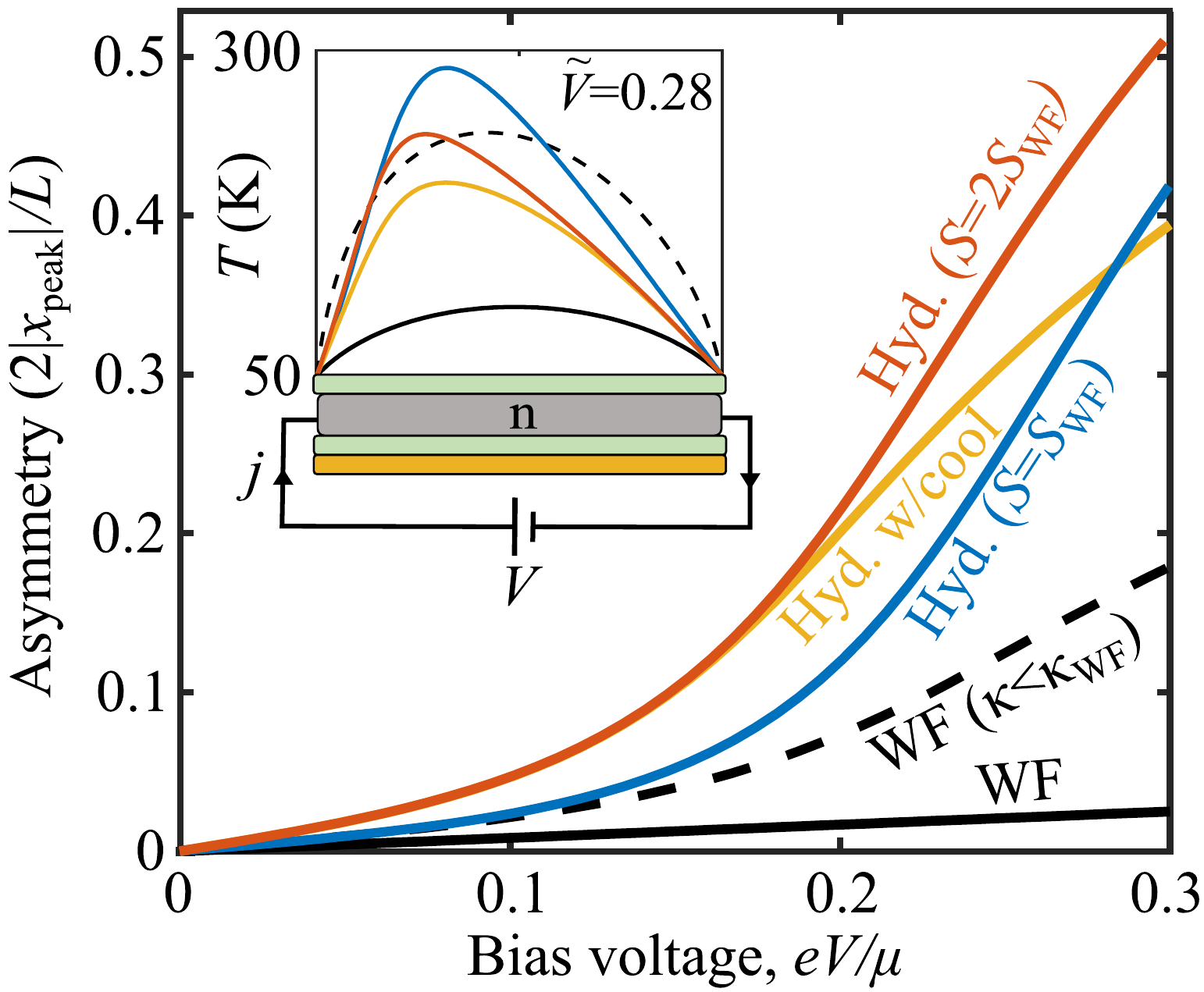}

\end{overpic}

\end{center}

\caption{(Color online)
E-e interactions facilitate extreme spatial asymmetry (Thomson effect), here shown as the (normalized) position of the temperature peak, plotted against $\tilde{V} = eV/\mu$. The five curves are the same as in Fig.~\ref{fig:two}. 
Inset: temperature profiles at $\tilde{V}=0.28$. ($L=5$ $\mu\rm m$, $\mu=0.1$ eV).
}

\label{fig:three}

\end{figure}

In the high-bias case, the Seebeck and Peltier terms skew the temperature profile in the direction of particle flow (electrons move left in the inset of Fig.~\ref{fig:three}). Such behavior is commonly referred to as the ``Thomson effect''~\cite{Bakan2014,Jungen2006,Grosse2014}, and can be understood as follows. The Seebeck effect produces an electric force that pushes particles in the direction of increasing temperature. Thus, on the upstream (downstream) side of the temperature peak, the Joule heating increases (decreases). This pushes the peak {\it upstream}. At the same time, heat is carried along with the particle flow (Peltier effect), and thus shifts the temperature peak {\it downstream}. The Thomson effect results from the competition between the two, and since the Peltier contribution is twice as large, the temperature profile is ultimately skewed downstream.


In the WF regime, one now finds:
\begin{equation} \label{eq:T_WF_approx_sol_high_bias}
f_{\rm J}(u;{\tilde V}) = \frac{2}{\tilde{V}}\frac{e^{\tilde{V}}(u-1)+2e^{\frac{1}{2}\tilde{V}(1-u)}-u-1}{1-e^{\tilde{V}}}
~.
\end{equation}
where $\tilde{V}=eV/\mu$. In the inset of Fig.~\ref{fig:three}, we plot the temperature profiles in the high-bias regime ($\tilde{V}=0.28$), which are numerically calculated in the hydrodynamic case (colored curves). In addition to the substantially increased amplitude, the hydrodynamic regime also exhibits far more spatial asymmetry. This is quantified in the main part of Fig.~\ref{fig:three} as the normalized position of the temperature peak, plotted against $\tilde{V}$. While the asymmetry is barely visible in the WF case, the peak shifts as much as 50$\%$ of the way to the contact in the hydrodynamic regime, even for the relatively modest $\tilde{V}$ considered here. With the use of spatially resolved temperature probes~\cite{Grosse2011,Halbertal2017,Freitag2010,Mecklenburg2015}, this anomalously large Thomson effect could potentially be a clear experimental signature of hydrodynamic heat transport. As in Figs.~\ref{fig:one}-\ref{fig:two}, we also display the WF case with a reduced $\kappa_{\rm WF}$ (black dashed). Clearly, its overall reduction is far from sufficient to produce the same level of asymmetry, indicating that the $1/T$-dependence is crucial. The reason is two-fold: First, the level of asymmetry is determined by the competition between the conduction cooling and Thomson terms. In the hydrodynamic regime, the former term becomes very weak near the peak, allowing for a stronger Thomson effect. Second, as pointed out in the PTE scenario, the $1/T$-dependence gives a more convex temperature profile, which further amplifies the Thomson term proportional to $T\partial_x T$. 

{\it Summary and conclusions.---}
We have here shown that e-e interactions can cause both a dramatic enhancement of TE efficiency and novel signatures, such as an anomalous Thomson effect, convex regions and superlinear temperature-power curves. Our findings offer new ways of experimentally observing hydrodynamic heat transport, and pave the way for the first TE applications of electron hydrodynamics. The latter would be improved even further if realized in materials with higher $S$.
\acknowledgments
We would like to thank Mikhail Lukin, Eugene Demler, Javier Sanchez-Yamagishi, Bo Dwyer, Jennifer Coulter, Giovanni Vignale and Mohammad Zarenia for helpful discussions.

\appendix

\begin{widetext}

\section{S1: Role of phonons}
When heat is initially injected into the electronic system, it must be transferred to the phonons before phononic heat transport can occur. We here consider heat loss mechanisms to phonons in graphene to show that the weak e-ph coupling bottlenecks this process in the regime studied in the main text. We also show that losses to hBN phonon-polaritons are more important, but that electronic conduction to the contacts is still the dominant cooling mechanism. 

We start out with the set of equations describing the electronic and phononic systems, depicted in Fig. $\rm{\ref{fig:1}}$:

\begin{equation} \label{eq:1}
\partial_x (\kappa_{\rm{e}}\partial_x T_{\rm{e}})-p_{\rm{e-ph}}^{\rm{hBN}}-p_{\rm{e-ph}}^{\rm{gr}}+p_{\rm{in}}(x)=0
\end{equation}
\begin{equation} \label{eq:2}
\partial_x (\kappa_{\rm{ph}}\partial_x T_{\rm{ph}})+p_{\rm{e-ph}}^{\rm{gr}}=0,
\end{equation}

where $p_{\rm{e-ph}}^{\rm{gr}}$ describes heat loss from electrons to graphene phonons, $p_{\rm{e-ph}}^{\rm{hBN}}$ describes heat loss to hBN phonon-polaritons ($p_{\rm{out}}$ in the main text), and $p_{\rm{in}}(x)$ is external laser heating. For simplicity, we have here considered one graphene phonon branch at a time, so $T_{\rm{ph}}$ is the temperature of the relevant phonon branch. 

\begin{figure}[H]

\begin{center}

\begin{overpic}[width=4.05in]{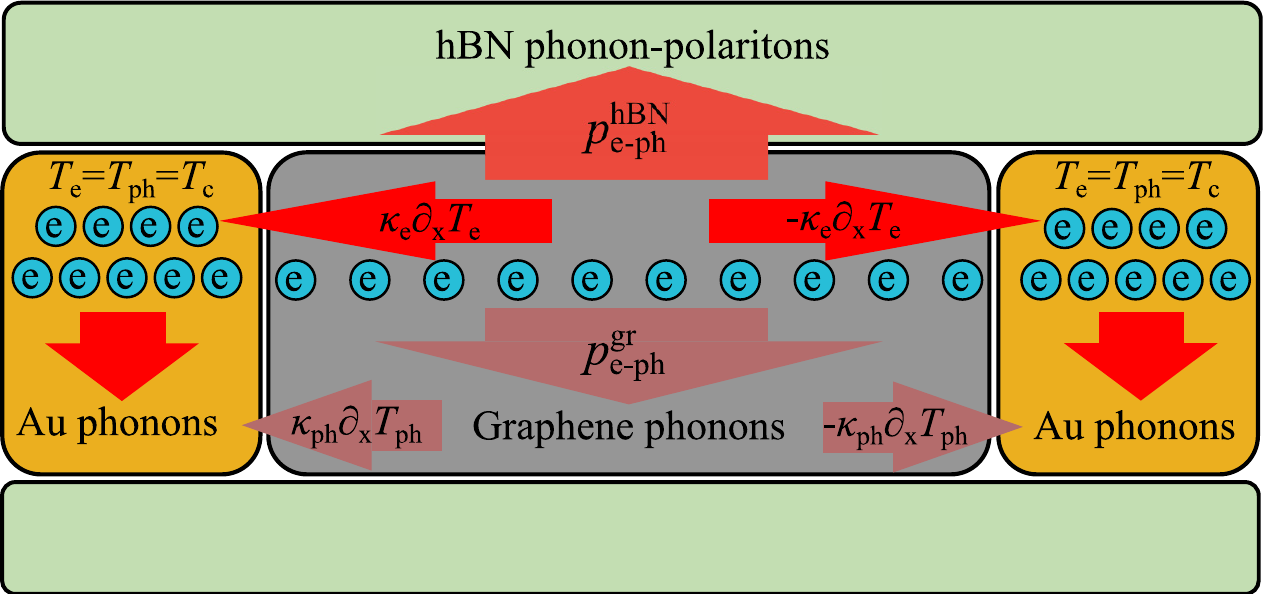}

\end{overpic}

\end{center}

\caption{Heat flow diagram of the encapsulated graphene system considered in the main text. The color intensity of the red arrows indicate the hierarchy of the different cooling channels (shown below). Due to the weak coupling to graphene phonons, only a very small portion ($<3\%$) of the heat is conducted by these phonon modes. More heat is transferred to hBN phonon-polaritons, and this cooling channel is therefore included in the main text. We note that electronic conduction to the contacts still dominates over all phonon channels in the regime considered in our work. The thick gold (Au) contacts act as thermal anchors ($T_{\rm{e}}(L/2)=T_{\rm{e}}(-L/2)=T_{\rm{c}}$) due to their larger electronic heat capacity and much stronger e-ph coupling than in graphene.  
}

\label{fig:1}

\end{figure}


The graphene phonon cooling mechanisms we shall consider are described in the table below:
\begin{center}
\begin{tabular}{| c| c | c| }
  \hline			
  \textbf{Mechanism} & \textbf{Expression for $\bm{p_{\rm{e-ph}}^{\rm{gr}}}$} & \textbf{Max. contribution} \\
    \hline 
  \begin{tabular}{@{}c@{}} Acoustic phonons,\\normal collisions (NC) \cite{Song2012} \end{tabular} & $\pi N (2\rho)^{-1}D^2\nu^2(\mu)\hbar k_{\rm{F}}^2k_{\rm{B}}(T_{\rm{e}}-T_{\rm{ph}})$ & 2.4\% \\
    \hline 
 \begin{tabular}{@{}c@{}} Acoustic phonons,\\supercollisions (SC) \cite{Song2012} \end{tabular}  & $9.62\frac{D^2\nu^2(\mu)k_{\rm{B}}^3}{2\rho s^2\hbar k_{\rm{F}}L_{\rm{mfp}}}\left(T_{\rm{e}}^3-T_{\rm{ph}}^3\right)$ 
& 0.22\% \\
  \hline  
   \begin{tabular}{@{}c@{}} $\rm{\:}$  \\Optical phonons \cite{Bistritzer2009}\\ $\rm{\:}$  \end{tabular} & $\frac{2\left(N_{\rm{e}}(E_{\rm{0}})-N_{\rm{ph}}(E_{\rm{0}})\right)N^2 \hbar^3 v_{\rm{F}}^2}{\rho a^4}{\displaystyle \int_{-\infty}^{\infty}} \nu(E)\nu(E-E_{\rm{0}})\left[f(E-E_{\rm{0}})-f(E)\right]dE$ & 0.07\% \\
  \hline
\end{tabular}
\end{center}

where $N=4$ is the number of spin/valley flavors in graphene, $\rho=7.61\cdotp10^{-7}\rm{\:kgm}^{-2}$ is the mass density of graphene, $D\approx20$ eV is the deformation potential, $\nu$ is the density of states per spin/valley flavor, $s\approx21$ km/s is the speed of sound, $L_{\rm{mfp}}$ is the mean free path, $N_{\rm{e(ph)}}$ is the Bose distribution function at the electron (phonon) temperature, $a=0.142$ nm is the C-C spacing in the graphene lattice, $E_{\rm{0}}\approx200$ meV is the optical phonon energy, and $f$ is the Fermi-Dirac distribution function. We note that the expressions for acoustic phonon cooling are exact only in the limit $T_{\rm{e}},T_{\rm{ph}}\gg T_{\rm{BG}}\equiv 2\hbar k_{\rm{F}}v_{\rm{s}}/k_{\rm{B}}$ ($T_{\rm{BG}}$ is the Bloch-Gruneisen temperature), but as shown in Ref. \cite{Hwang2008}, they are almost perfect approximations in our case, where $T_{\rm{e}}\geq T_{\rm{ph}}\sim T_{\rm{BG}}=49$ K. As indicated in Eq. ($\rm{\ref{eq:1}}$), we shall also include losses to phonon-polaritons in hBN. 

Since we reach similar electronic temperatures in the laser- and Joule heating scenarios, we will here consider only the former ($p_{\rm{in}}(x)=P/W\cdotp\delta(x)$). Now, to confidently show that graphene phonons do not play a dominant role, we shall consider the worst case, i.e. the one that maximizes heat loss to phonons. Since $p_{\rm{e-ph}}^{\rm{gr}}$ is a decreasing function of $T_{\rm{ph}}$, we consider the case where the graphene phonons conduct the heat to the contacts instantaneously ($\kappa_{\rm{ph}}=\infty$) and thus $T_{\rm{ph}} (x)=T_{\rm{c}}$ everywhere.

In order to compare the contributions of electronic conduction and heat loss to phonons, we integrate Eq. ($\rm{\ref{eq:1}}$) to obtain:
\begin{equation} \label{eq:3}
\overbrace{P/W}^{\rm{Total\:laser\:power\:in}}=\overbrace{-2\kappa_{\rm{e}}(\partial_x T_{\rm{e}})_{x=L/2}}^{\rm{Electronic\:conduction\:to\:contacts}}+\overbrace{\int_{-L/2}^{L/2}p_{\rm{e-ph}}^{\rm{hBN}}dx}^{\rm{Heat\:loss\:to\:hBN\:phonons}}+\overbrace{\int_{-L/2}^{L/2}p_{\rm{e-ph}}^{\rm{gr}}dx}^{\rm{Heat\:loss\:to\:gr.\:phonons}}
~.
\end{equation}
which is merely a statement of energy conservation of the 1D heat flows. Fig. $\rm{\ref{fig:2}}$ shows the contributions from the different cooling mechanisms as a function of laser power, for the same parameters as used in the main text. It is immediately evident that the graphene phonons do not play an important role. 

\begin{figure}[H]

\begin{center}

\begin{overpic}[width=4.57in]{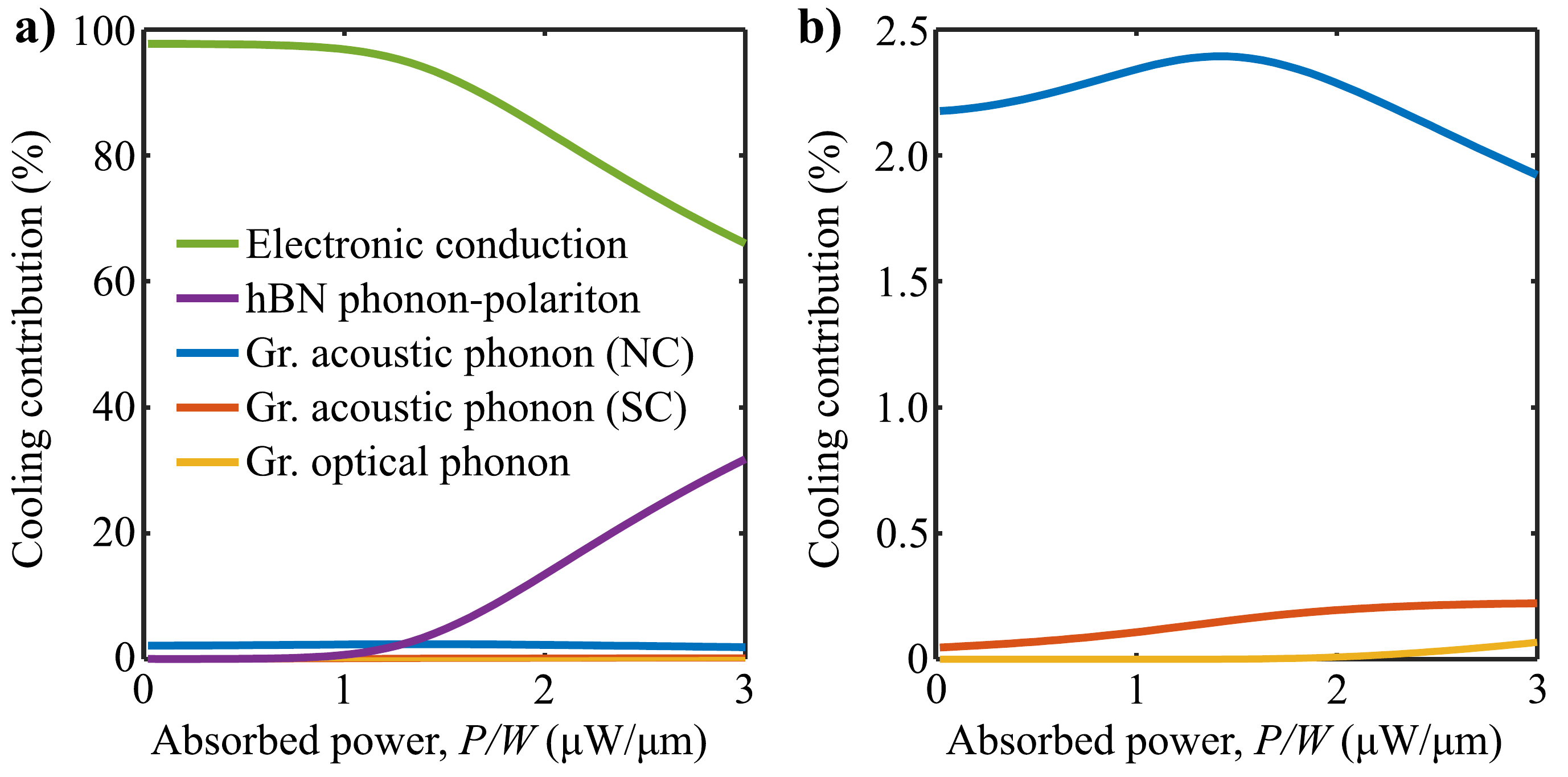}

\end{overpic}

\end{center}

\caption{(a) The contributions of various cooling channels as a function of laser power, for the parameters used in the main text. For these values of $P/W$, electronic conduction to the contacts (green) is the dominant cooling channel. Out of the phononic channels, heat loss to phonon-polaritons in the hBN encapsulant (purple) dominates over cooling through graphene phonons. (b) Zoomed-in version of (a) to highlight the low heat loss to graphene phonons.
}

\label{fig:2}

\end{figure}


\section{S2: Device length dependence}
We here comment on the length dependence of the results from the photoenergy harvesting scenario presented in the main text. The roles of electronic conduction to the contacts and phonon cooling depend on the length of the device. While phonon cooling increases with device length due to increased area for electron-phonon heat transfer, electronic conduction has the opposite behavior. Its ability to cool the system decreases in longer devices, since the heat sinks (contacts) are further away from the heating spot. This is immediately seen from the fact that $\theta\propto L$ in the temperature profiles, $T_{\rm{Hyd}} (x)=T_{\rm c}\sqrt{\left(1+\gamma^{-1}\right)\exp\left[2\theta \gamma f_{\rm PTE}(x)\right]-\gamma^{-1}}$ and $T_{\rm{WF}}(x)=T_{\rm{0}} \sqrt{1+2\theta f_{\rm{PTE}}(x)}$. Thus, heat loss to phonons (here hBN phonon-polaritons) becomes more important in longer devices, as shown in Fig. $\rm{\ref{fig:3}}$(a). In our work, we have therefore chosen a relatively short, but experimentally realistic device length ($L=5$ $\mu$m). 

Due to the points presented above, the device length also affects the thermoelectric efficiency (Fig. $\rm{\ref{fig:3}}$(b)). Since the electronic and phononic cooling mechanisms have opposite dependence on $L$, the efficiency depends non-monotonically on device length and is optimized near the $L$ that makes their contributions equal. 
We note that a large portion of the heat transferred to the hBN phonon-polaritons has first been (electronically) conducted through parts of the device, and thus still ``imprints'' the hydrodynamic form of $\kappa_{\rm{e}}$ on the temperature profile.
 
\begin{figure}[H]

\begin{center}

\begin{overpic}[width=3.83in]{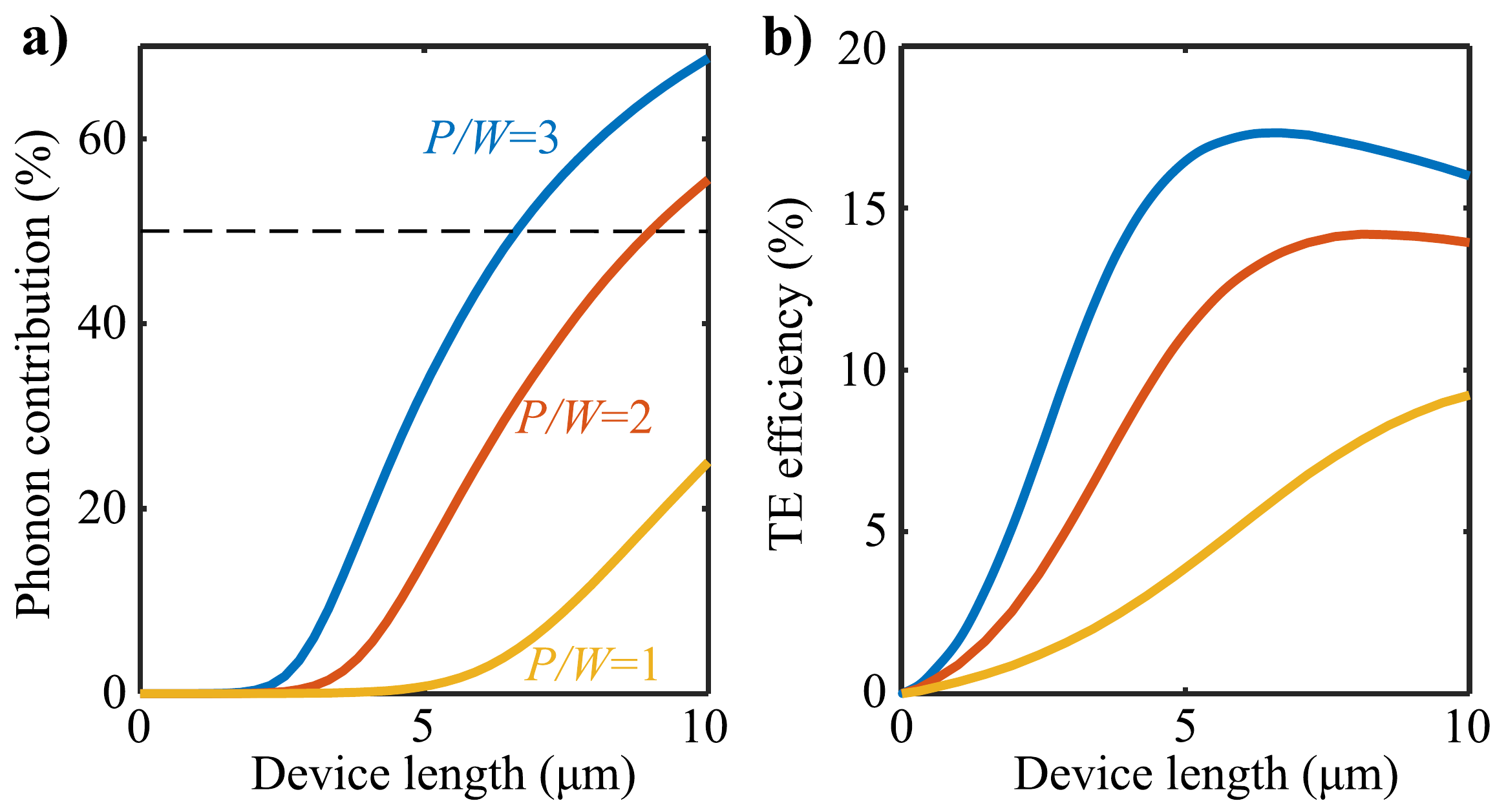}

\end{overpic}

\end{center}

\caption{(a) Relative amount of heat that goes to hBN phonon-polaritons as a function of device length, for laser powers $P/W=1$, 2 and 3 $\mu$W/$\mu$m. Dashed line indicates where electronic and phononic cooling contributions are equal (50\%). (b) Thermoelectric efficiency as a function of device length for same laser powers as in (a). The efficiency is observed to be optimized near the length that gives equal electron and phonon contributions. Other parameters are the same as in the main text. 
}

\label{fig:3}

\end{figure}

Finally, we also comment on the length-dependent role of phonons in the more general hydrodynamic case, by considering phonon cooling on the common form, $p_{\rm{e-ph}}=\Sigma\left(T_{\rm{e}}^{\delta}-T_{\rm{ph}}^{\delta}\right)$\cite{Balkan1995,Ridley1991,Kaasbjerg2014}. 

We evaluate the phonon contribution in Eq. ($\rm{\ref{eq:3}}$) by using $T_{\rm{e}}(x)=T_{\rm{Hyd}}(x)$ (Eq. (5) in the main text). In doing so, we neglect the effects of $p_{\rm{e-ph}}$ on the electronic temperature, and thus slightly overestimate $T_{\rm{e}}(x)$ and $p_{\rm{e-ph}}$. In Figure $\rm{\ref{fig:4}}$ (a)-(c), we show the (power-dependent) device length for which $20\%$ of the injected heat goes to the phononic system for various values of $\Sigma$, and $\delta=$1, 3 and 5. These are commonly encountered values of $\delta$: $\delta=1$ corresponds to acoustic phonon mechanisms in most materials in the equipartition regime ($T_{\rm{e}},T_{\rm{ph}}\gg T_{\rm{BG}}$) \cite{Kaasbjerg2014}, and $\delta=3$ and 5 represent, respectively, acoustic phonon cooling in 2DEGs \cite{Balkan1995} and 3D metals \cite{Wellstood1994} in the Bloch-Gruneisen regime ($T_{\rm{e}},T_{\rm{ph}}\ll T_{\rm{BG}}$). Aside from the cooling power, we have here used the same parameters ($T_{\rm{int}}$ and $\sigma$) as those used in the main text. As in the graphene-specific case, lower laser powers allow electronic conduction to remain dominant in longer devices. 

\begin{figure}[H]

\begin{center}

\begin{overpic}[width=3.92in]{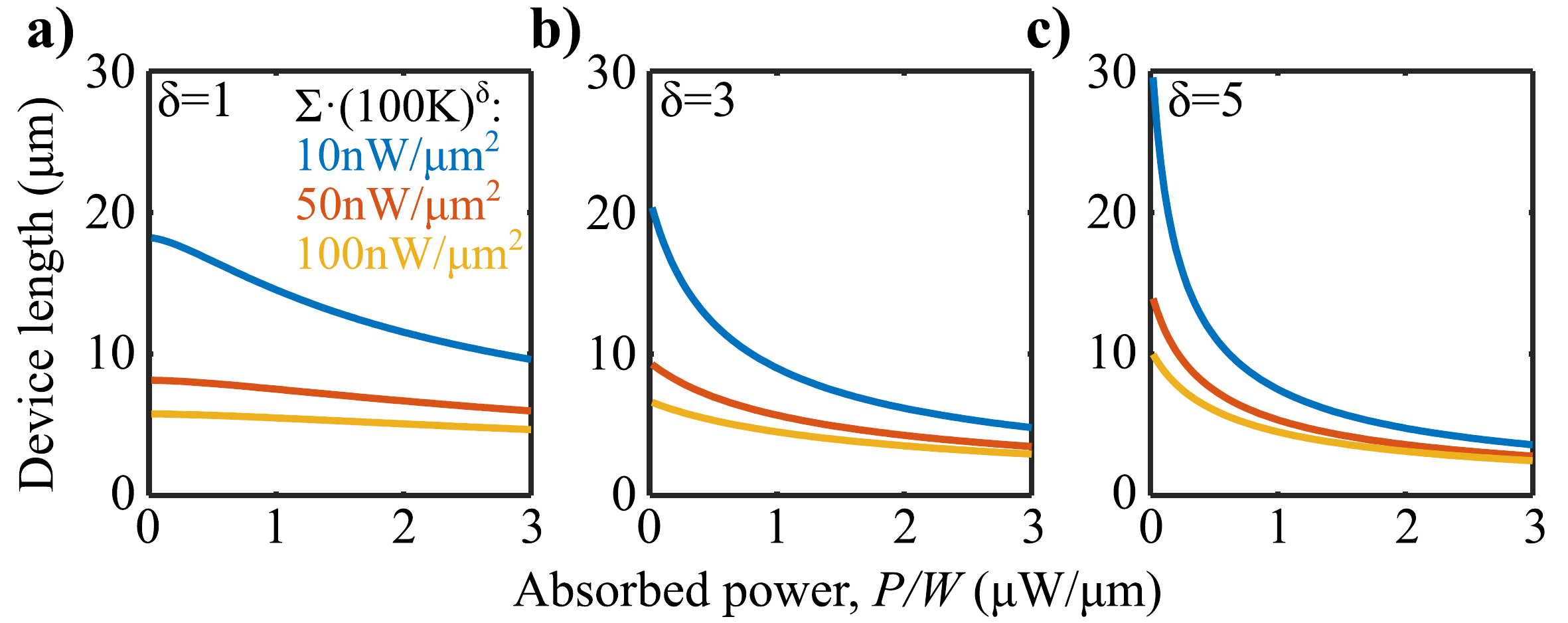}

\end{overpic}

\end{center}

\caption{
(a)-(c) Device length for which 20\% of the injected heat enters the graphene phonon system, as a function of laser power. This is evaluated for $\Sigma\cdotp (\rm{100\:K})^{\delta}=10,\rm{\:}50\rm{\:and\:}100\rm{\:nW/\mu m^2}$ (blue, red and yellow, respectively), and $\delta=1\rm{,\:}3$ and $5$ in (a), (b) and (c), respectively. Other parameters are the same as in the main text. 
}

\label{fig:4}

\end{figure}

\end{widetext}

\bibliography{biblio_arxiv}

\end{document}